\documentclass[prd,twocolumn,showpacs,nofootinbib]{revtex4}
\usepackage{graphicx}
\usepackage{dcolumn}
\usepackage{amssymb}
\usepackage{mathrsfs}
\usepackage{amsmath}
\usepackage{epsfig}
\usepackage[dvips]{color}
\usepackage{hhline}

\begin{document}

\title{Quasi-degenerate dark photon and dark matter}

\author{Hang Zhou}
\email{zhouhang@cczu.edu.cn}
\affiliation{School of Microelectronics and Control Engineering, Changzhou University, Changzhou, 213164, China}

\begin{abstract}
We introduce an $SU(2)_X^{}$ dark sector without any fermions and then realize a non-abelian kinetic mixing between the dark $SU(2)_X^{}$ gauge fields and the standard model $SU(2)_L^{}\times U(1)_Y^{}$ gauge fields. While one of the dark gauge bosons becomes a dark photon, the others can keep stable to form a dark matter particle. The nearly degenerate masses of dark photon and dark matter could be tested if the dark photon and the dark matter are both observed in the future.
\end{abstract}


\maketitle

\section{Introduction}

The idea of dark photon \cite{okun1982,holdom1986,fh1991} has been studied by many theorists and experimentalists. Usually, the interactions of the dark photon to the visible matter are assumed from an abelian kinetic mixing between the standard model (SM) $U(1)_Y^{}$ gauge group and a dark $U(1)_X^{}$ gauge group. This $U(1)$ kinetic mixing can not be sizable due to the constraints from different experiments including low energy colliders \cite{Merkel:2014avp,Lees:2014xha,Kou:2018nap}, meson decays \cite{Pospelov:2008zw,Batley:2015lha,Banerjee:2018vgk}, beam dump experiments \cite{Riordan:1987aw,Blumlein:2013cua,Blumlein:2011mv} and high-energy colliders \cite{Curtin:2014cca,Anelli:2015pba,Ilten:2016tkc,Blinov:2017dtk,Aaij:2019bvg,Sirunyan:2019wqq}. However, it is not obvious to explain the smallness of a renormalizable $U(1)$ kinetic mixing.

In this paper we shall consider the non-abelian kinetic mixing to realize another possibility that the dark photon originates from a dark $SU(2)_X^{}$ gauge group and hence its couplings to the matter do not appear at renormalizable level \cite{ccf2009,bcn2016,ahopr2017}. In the presence of the non-abelian kinetic mixing between the dark $SU(2)_X^{}$ gauge group and the SM $SU(2)_L^{}\times U(1)_Y^{}$ gauge groups, one of the dark gauge bosons becomes a dark photon, while the others keep stable to form a dark matter particle. This scenario predicts a nearly degenerate mass spectrum of dark photon and dark matter.

\section{Dark sector}

Besides the dark gauge fields $X^{1,2,3}_\mu$, the $SU(2)_X^{}$ dark sector only contains a dark Higgs doublet $\chi=(\chi_1^{},\chi_2^{})^T_{}$. The full Lagrangian of the dark sector is
\begin{eqnarray}
\mathcal{L}_{\textrm{Dark}}^{} &=& -\frac{1}{4} X_{\mu\nu}^{a} X_{}^{a\mu\nu} + \left(D_\mu^{}\chi \right)^\dagger_{}  \left(D^\mu_{}\chi \right) - \mu_\chi^2 \chi^\dagger_{}\chi \nonumber\\
&&- \lambda_\chi^{} \left(\chi^\dagger_{}\chi\right)^2_{} ~~\textrm{with}\nonumber\\
&& X_{\mu\nu}^a = \partial_\mu^{} X_\nu^{a} - \partial_\nu^{} X_\mu^{a} + i g_X^{} \varepsilon^{abc}_{} X^{b}_{\mu} X^{c}_{\nu}\,,\nonumber\\
&&D_\mu^{} \chi =  \left(\partial_\mu^{} - i g_X^{} \frac{\tau_a^{}}{2} X^a_\mu\right) \chi\,,
\end{eqnarray}
where $g_X^{}$ is the $SU(2)_X^{}$ gauge coupling. At the renormalizable level, the dark sector can interact with the SM only through the Higgs portal as below,
\begin{eqnarray}
\mathcal{L}_{\textrm{int}}^{} &=&-\lambda_{\phi\chi}^{} \phi^\dagger_{}\phi \chi^\dagger_{}\chi\,.
\end{eqnarray}
Here $\phi=(\phi_{}^{0},\phi_{}^{-})^T_{}$ is the SM Higgs doublet for the spontaneous electroweak symmetry breaking, i.e.
\begin{eqnarray}
\phi \Rightarrow \phi =\left[\begin{array}{c} \frac{1}{\sqrt{2}}\left(v_\phi^{} + h_\phi^{}\right)\\
[2mm]
0
\end{array}\right] \,.
\end{eqnarray}

When the dark Higgs doublet $\chi$ develops its vacuum expectation value (VEV) for spontaneously breaking the $SU(2)_X^{}$ symmetry, i.e.
\begin{eqnarray}
\chi \Rightarrow \chi=\left[\begin{array}{c} \frac{1}{\sqrt{2}}\left(v_\chi^{} + h_\chi^{}\right)\\
[2mm]
0
\end{array}\right] \,,
\end{eqnarray}
the dark gauge fields $X^{1,2,3}_\mu$ can obtain their masses,
\begin{eqnarray}
\mathcal{L}_{\textrm{Dark}}^{} &\supset&\frac{1}{8} g_X^2 v_\chi^2 \sum_{a=1,2,3}^{}X^{a}_{\mu} X^{a\mu}_{} \nonumber\\
&=&m_X^2 X^{+}_\mu X^{-\mu}_{}  + \frac{1}{2} m_X^2 X^{3}_{\mu} X^{3\mu}_{}~~\textrm{with}\nonumber\\
&&X^{\pm}_\mu = \frac{1}{\sqrt{2}}\left(X^{1}_\mu \mp i X^{2}_\mu\right)\,,~~m_X^2 = \frac{g_X^2   v_\chi^2 }{4} \,.
\end{eqnarray}
In the following, we shall conveniently refer to $X^{\pm}_\mu$ as the charged dark gauge bosons and $X^{3}_\mu$ as the neutral dark gauge boson although these dark gauge bosons do not carry the ordinary electric charge.

\section{Non--abelian kinetic mixing}

So far the dark gauge bosons $X^{\pm}_\mu$ and $X^{3}_\mu$ are exactly degenerate. This feature can be modified if the dark $SU(2)_X^{}$ gauge fields $X^{1,2,3}_{\mu}$ have a non-abelian kinetic mixing with the SM $SU(2)_L^{}\times U(1)_Y^{}$ gauge fields $W^{1,2,3}_\mu, B^{}_{\mu}$, i.e.
\begin{eqnarray}
\label{effective}
\mathcal{L} &\supset& -\frac{1}{\Lambda^2_{6}} \chi^\dagger_{} \frac{\tau_a^{}}{2} \chi X^{a}_{\mu\nu} B^{\mu\nu}_{}-\frac{1}{\Lambda^4_{8}} \chi^\dagger_{} \frac{\tau_a^{}}{2} \chi X^{a}_{\mu\nu} \phi^\dagger_{} \frac{\tau_b^{}}{2} \phi W^{b\mu\nu}_{}\nonumber\\
&&\textrm{with}~~  B^{}_{\mu\nu}=\partial_\mu^{} B_\nu^{} - \partial_\nu^{} B_\mu^{}\,,\nonumber\\
&&\quad\quad \quad \!W^{a}_{\mu\nu}=\partial_\mu^{} W_\nu^{a}
- \partial_\nu^{} W_\mu^{a} +  i g \varepsilon^{abc}_{} W^{b}_\mu W^{c}_\nu\,.
\end{eqnarray}
Here $g$ is the $SU(2)_L^{}$ gauge coupling. As for the other non-abelian kinetic mixing, they can be forbidden by an additional $U(1)_D^{}$ global symmetry under which the dark Higgs doublet $\chi$ is non-trivial, i.e.
\begin{eqnarray}
\mathcal{L} ~/ \!\!\!\!\!\supset -\frac{1}{\tilde{\Lambda}^2_{6}} \chi^\dagger_{} \frac{\tau_a^{}}{2} \tilde{\chi} X^{a}_{\mu\nu} B^{\mu\nu}_{}-\frac{1}{\tilde{\Lambda}^4_{8}} \chi^\dagger_{} \frac{\tau_a^{}}{2} \tilde{\chi} X^{a}_{\mu\nu} \phi^\dagger_{} \frac{\tau_b^{}}{2} \phi W^{b\mu\nu}_{}\,.
\end{eqnarray}

The non-abelian kinetic mixing (\ref{effective}) can lead to the kinetic mixing between the dark $X^{3}_\mu$ and the SM $W^{3}_\mu,B_\mu^{}$, i.e.
\begin{eqnarray}
\mathcal{L}&\supset & -\frac{\epsilon_B^{}}{2}  \tilde{X}^{3}_{\mu\nu} B^{\mu\nu}_{}-\frac{\epsilon_W^{}}{2}  \tilde{X}^{3}_{\mu\nu} \tilde{W}^{3\mu\nu}_{}~~\textrm{with}\nonumber\\
&&\epsilon_B^{} \equiv \frac{v_\chi^2 }{2\Lambda^2_{6}}\,,~~ \tilde{X}^{3}_{\mu\nu}=\partial_\mu^{} X_\nu^{3} - \partial_\nu^{} X_\mu^{3}\,,\nonumber\\
&&\epsilon_W^{}  \equiv \frac{v_\chi^2 v_\phi^2}{8\Lambda^4_{8}}\,,~~ \tilde{W}^{3}_{\mu\nu}=\partial_\mu^{} W_\nu^{3} - \partial_\nu^{} W_\mu^{3}\,,
\end{eqnarray}
after the dark and electroweak symmetry breaking, i.e.
\begin{eqnarray}
SU(2)_X^{}\times U(1)_D^{} &\stackrel{\langle\chi\rangle}{\longrightarrow}& U(1)_{DE}^{}\,,\\
SU(2)_L^{}\times U(1)_Y^{}&\stackrel{\langle\phi\rangle}{\longrightarrow}& U(1)_{em}^{}\,.
\end{eqnarray}
Here $U(1)_{DE}^{}$ is an unbroken global symmetry to define a dark electric charge. Therefore, the neutral dark gauge boson $X^{3}_\mu$ indeed becomes a dark photon so that it can be distinguished from the charged dark gauge bosons $X^{\pm}_{\mu}$. More details will be given shortly.

The effective operators in Eq. (\ref{effective}) can be induced by integrating out certain scalar(s) crossing the dark and SM sectors. For example, an $[SU(2)_X^{}]$-doublet scalar with $U(1)_Y^{}$ charge can mediate the $SU(2)_X^{} \times U(1)_Y^{}$ mixing, an $[SU(2)_X^{}]$-doublet and $[SU(2)_L^{}]$-triplet scalar without $U(1)_Y^{}$ charge can mediate the $SU(2)_X^{} \times SU(2)_L^{}$ mixing, while an $[SU(2)_X^{}\times SU(2)_L^{}]$-bidoublet scalar with $U(1)_Y^{}$ charge or an $[SU(2)_X^{}]$-doublet and $[SU(2)_L^{}]$-triplet with $U(1)_Y^{}$ charge can mediate both of the $SU(2)_X^{} \times U(1)_Y^{}$ and $SU(2)_X^{} \times SU(2)_L^{}$ mixing. It should be noted these crossing scalars are also non-trivial under the $U(1)_D^{}$ global symmetry to forbid some unexpected couplings.

\subsection{The case only with $SU(2)_X^{}\times U(1)_Y^{}$ kinetic mixing }

In this case, the crossing scalar could be an $[SU(2)_X^{}]$-doublet scalar with a non-trivial $U(1)_Y^{}$ hypercharge,
\begin{eqnarray}
\eta=\left[\begin{array}{c} \eta_1^{Y_\eta^{}}\\
[3mm]
\eta_2^{Y_\eta^{}}
\end{array}\right] \,,
\end{eqnarray}
with $Y_\eta^{}$ being the $U(1)_Y^{}$ hypercharge. The crossing scalar $\eta$ also carries a $U(1)_D^{}$ charge as the same with the dark Higgs doublet $\chi$. The following $\chi-\eta$ couplings,
\begin{eqnarray}
\mathcal{L}\supset  -\lambda^{}_{1} \chi^\dagger_{}\eta \eta^\dagger_{}\chi - \lambda^{}_{2} \tilde{\chi}^\dagger_{}\eta \eta^\dagger_{}\tilde{\chi}   \,,
\end{eqnarray}
then can break the mass degeneracy between the $\eta_{1,2}^{}$ components, i.e.
\begin{eqnarray}
\mathcal{L}&\supset & -m_{\eta_1^{}}^2 \eta^{\ast}_{1}\eta^{}_{1}   -m_{\eta_2^{}}^2 \eta^{\ast}_{2}\eta^{}_{2}  ~~\textrm{with}\nonumber\\
&&
m_{\eta_1^{}}^2  - m_{\eta_2^{}}^2 = \frac{1}{2}\left(\lambda_{1}^{} - \lambda_{2}^{}\right) v_\chi^2\,.
\end{eqnarray}
We then can compute the $X^{3}_{}-B$ kinetic mixing at one-loop level \cite{fltw2011},
\begin{eqnarray}
\label{kinetic}
\epsilon_B^{} =  \frac{g_X^{} g' Y_\eta^{} }{96\pi^2_{}}   \ln \left(\frac{m_{\eta_1^{}}^2}{m_{\eta_2^{}}^2} \right)\,,
\end{eqnarray}
with $g'$ being the $U(1)_Y^{}$ gauge coupling.

We should keep in mind the crossing scalar $\eta$ carries a non-zero electric charge so that it should be heavy enough and decay before the BBN. Otherwise, it should have been ruled out experimentally. This means we need other $[SU(2)_X^{}]$-singlet mediator scalar(s) to make the crossing scalar $\eta$ unstable. For example, the mediator scalar can be a singly charged dilepton scalar $\xi$ with the following couplings,
\begin{eqnarray}
\mathcal{L} \supset  - f_\xi^{}\xi \bar{l}_L^c i\tau_2^{} l_L^{} -\lambda_{\xi\eta\chi}^{} \xi^2_{}\eta^\dagger_{}\chi +\textrm{H.c.}\,.
\end{eqnarray}
We can also consider a doubly charged dilepton scalar $\zeta$ to be the mediator scalar. The related couplings include
\begin{eqnarray}
\mathcal{L} \supset - f_\zeta^{} \zeta \bar{e}_R^c e_R^{}  -\mu_{\zeta\eta\chi}^{} \zeta \eta^\dagger_{}\chi  +\textrm{H.c.}\,.
\end{eqnarray}
In this case, the formula (\ref{kinetic}) should be modified because of the $\eta_{1,2}^{}-\zeta$ mixing. Alternatively, the crossing scalar $\eta$ could be colored when the mediator is composed of certain diquark or leptoquark scalar(s). These mediator scalars may be useful in the generation of radiative neutrino masses \cite{zee1986,babu1988,bl2001}.

We would like to emphasise that the crossing scalars do not have the following gauge-invariant terms because of the $U(1)_D^{}$ global symmetry,
\begin{eqnarray}
\label{stable}
\mathcal{L} ~/\!\!\!\!\!\supset   - \lambda^{}_{12} \chi^\dagger_{}\eta \eta^\dagger_{}\tilde{\chi} - \tilde{\lambda}_{\xi\eta\chi}^{} \xi^2_{}\eta^\dagger_{}\tilde{\chi}- \tilde{\lambda}_{\zeta\eta\chi}^{} \zeta \eta^\dagger_{}\tilde{\chi} +\textrm{H.c.}\,.
\end{eqnarray}
In the absence of the above couplings, the $\eta^{}_{2}$ component of the crossing scalar $\eta$ can only decay into a charged dark gauge boson $X^{\pm}_{\mu}$ with a real or virtual $\eta^{}_{1}$ component. Therefore, the charged dark gauge bosons $X^{\pm}_{\mu}$ can keep stable.

\subsection{The case only with $SU(2)_X^{} \times SU(2)_L^{}$ kinetic mixing}

In this case, the crossing scalar could be an $[SU(2)_X^{}]$-doublet and $[SU(2)_L^{}]$-triplet scalar without $U(1)_Y^{}$ charge,
\begin{eqnarray}
\Delta =\left[\begin{array}{c}\delta_1^{}\\
[2mm]
\delta_2^{} \end{array}\right]~~\textrm{with}~~\delta_{i}^{}=\left[\begin{array}{cc}\frac{1}{\sqrt{2}}\delta^{0}_{i}&\delta^{+}_{i2}\\
[2mm]
\delta^{-}_{i1}& -\frac{1}{\sqrt{2}}\delta^{0}_{i} \end{array}\right]\,.
\end{eqnarray}
For simplicity we do not explicitly calculate the $X^3_{}-W^3_{}$ kinetic mixing, which should depend on the mass difference among the components of the crossing scalar $\Delta$, like the formula (\ref{kinetic}).

Note the crossing scalar $\Delta$ can decay through the following quartic coupling,
\begin{eqnarray}
\mathcal{L}\supset  -\lambda_{\chi \phi \Delta}^{}\left[\tilde{\phi}^T_{} i\tau_2^{}   \left( \chi^\dagger_{}  \Delta\right)  \phi  +\textrm{H.c.}\right]\,,
\end{eqnarray}
without resorting to additional mediator scalars. Clearly, this crossing scalar can pick up a VEV $\langle\Delta\rangle$ after the dark and electroweak symmetry breaking.

\begin{eqnarray}
\langle\Delta\rangle =\left[\begin{array}{c}\langle\delta_1^{}\rangle\\
[2mm]
0 \end{array}\right]~~\textrm{with}~~\langle\delta_{1}^{}\rangle=\left[\begin{array}{cc}\frac{1}{\sqrt{2}}\langle\delta^{0}_{1}\rangle&0\\
[2mm]
0& -\frac{1}{\sqrt{2}}\langle\delta^{0}_{1}\rangle \end{array}\right]\,.
\end{eqnarray}

\subsection{The case with both $SU(2)_X^{} \times SU(2)_L^{}$ and $SU(2)_X^{} \times U(1)_Y^{}$ kinetic mixing}

In this case, the crossing scalar could be an $[SU(2)_X^{}\times SU(2)_L^{}]$-bidoublet scalar with $U(1)_Y^{}$ charge,
\begin{eqnarray}
\Sigma =\left[\begin{array}{c}\sigma_1^{}\\
[2mm]
\sigma_2^{} \end{array}\right]~~\textrm{with}~~\sigma_{i}^{}=\left[\begin{array}{c} \sigma^{0}_{i}\\
[2mm]
\sigma^{-}_{i} \end{array}\right]\,.
\end{eqnarray}
or an $[SU(2)_X^{}]$-doublet and $[SU(2)_L^{}]$-triplet with $U(1)_Y^{}$ charge,
\begin{eqnarray}
\Omega =\left[\begin{array}{c}\omega_1^{}\\
[2mm]
\omega_2^{} \end{array}\right]~~\textrm{with}~~\omega_{i}^{}=\left[\begin{array}{cc}\frac{1}{\sqrt{2}}\omega^{+}_{i}&\omega^{++}_{i}\\
[2mm]
\omega^{0}_{i}& -\frac{1}{\sqrt{2}}\omega^{+}_{i} \end{array}\right]\,.
\end{eqnarray}
The formula on the $X^3_{}-B$ and $X^3_{}-W^3_{}$ kinetic mixing should be similar to Eq. (\ref{kinetic}). For simplicity we do not show the calculations.

The crossing scalars $\Sigma$ and $\Omega$ have the following couplings to the dark Higgs doublet $\chi$ and the SM Higgs doublet $\phi$, i.e.
\begin{eqnarray}
\mathcal{L}&\supset&  - \mu_{\chi \phi \Sigma}^{} \left( \chi^\dagger_{} \Sigma \phi  +\textrm{H.c.} \right)  \nonumber\\
&&- \lambda_{\chi \phi \Sigma}^{} \left[\phi^T_{}i \tau_2^{} \left(\chi^\dagger_{} \Omega\right) \phi  +\textrm{H.c.} \right]  \,.
\end{eqnarray}
Therefore, we need not introduce additional mediator scalars in this case.

\section{Dark photon}

We can remove the loop-induced kinetic mixing between the neutral dark gauge field $X^{3}_\mu$ and the SM gauge fields $B_\mu^{},W^{3}_\mu$ by a non-orthogonal rotation as below,
 \begin{eqnarray}
X_\mu^{3}&=&\frac{1}{\sqrt{1-\epsilon^2_{B}- \epsilon^2_{W}}} X'^{3}_\mu\,,\nonumber\\
B_\mu^{}&=&B'^{}_\mu - \frac{\epsilon_B^{}}{\sqrt{1-\epsilon^2_{B}  -\epsilon^2_{W}  }} X'^{3}_\mu\,,\nonumber\\
W_\mu^{3}&=&W'^{3}_\mu - \frac{\epsilon_W^{}}{\sqrt{1    - \epsilon^2_{B}   -\epsilon^2_{W}}} X'^{3}_\mu\,.
\end{eqnarray}
In the basis of $B'^{}_\mu, W'^{3}_\mu, X'^{3}_\mu$, we can obtain the massless photon and the $Z$ boson by
\begin{eqnarray}
A_\mu^{} &=& W'^3_\mu \sin \theta_W^{} + B'^{}_\mu \cos \theta_W^{}\,,\nonumber\\
Z_\mu^{} & =& W'^3_\mu \cos\theta_W^{}  - B'^{}_\mu \sin\theta_W^{}\,,\end{eqnarray}
where $\theta_W^{}$ is the Weinberg angle as usual, i.e. $\tan \theta_W^{} = g'/g $. For the following demonstration, we would like to conveniently denote the parameters,
\begin{eqnarray}
\epsilon_Z^{}&=&\epsilon_W^{} \cos\theta_W^{}   -  \epsilon_B^{} \sin\theta_W^{}   \,,\nonumber\\
 \epsilon_A^{} &=&  \epsilon_W^{} \sin\theta_W^{}   +  \epsilon_B^{} \cos\theta_W^{} \,,\nonumber\\
 \bar{\epsilon}_Z^{} &=& \frac{\epsilon_Z^{}}{\sqrt{1    - \epsilon^2_{Z}   -\epsilon^2_{A}}}\,,~~ \bar{\epsilon}_A^{} =\frac{\epsilon_A^{}}{\sqrt{1    - \epsilon^2_{Z}   -\epsilon^2_{A}}}  \,.
\end{eqnarray}

Now the $Z$ boson has a mass mixing with the $X'^3_{}$ boson, i.e.
\begin{eqnarray}
\mathcal{L} &\supset&  \frac{g^2_{} v_\phi^2 }{8 \cos^2_{}\theta_W^{}}\left(Z_\mu^{} -   \bar{\epsilon}_Z^{}   A'^{}_\mu \right) \left(Z^\mu_{} -   \bar{\epsilon}_Z^{} A'^\mu_{}\right) \nonumber\\
&&+ \frac{g_X^2  v_\chi^2}{8(1-\epsilon^2_{Z}- \epsilon^2_{A})}  A'^{}_{\mu} A'^{\mu}_{} \nonumber\\
&=&\frac{1}{2}m_Z^2 Z_\mu^{} Z^\mu_{} + \frac{1}{2} m_{A'}^2 A'^{}_\mu A'^\mu_{} + m_{ZA'}^2 Z_\mu^{} A'^\mu_{}~~\textrm{with}\nonumber\\
&&m_Z^2 = \frac{g^2_{} v_\phi^2 }{4 \cos^2_{}\theta_W^{}} \,,\nonumber\\
&&m_{A'}^2 =  \frac{g_X^2  v_\chi^2}{4(1-\epsilon^2_{Z}- \epsilon^2_{A})}+ \frac{ g^2_{} v_\phi^2 \epsilon_Z^2}{4 \cos^2_{}\theta_W^{}  \left(1    - \epsilon^2_{Z}   -\epsilon^2_{A}\right)}\,,\nonumber\\
&&m_{ZA'}^2= - \frac{ g^2_{} v_\phi^2  \epsilon_Z^{} }{4 \cos^2_{}\theta_W^{} \sqrt{1    - \epsilon^2_{Z}   -\epsilon^2_{A}}} \,.
\end{eqnarray}
Here we have identified the $X'^3$ boson as the $A'$ boson. The mass eigenstates of the $Z$ and $A'$ bosons should be
\begin{eqnarray}
\hat{Z}_\mu^{} &=&Z_\mu^{} \cos \zeta - A'^{}_\mu \sin \zeta  ~~\textrm{with}\nonumber\\
&&m^2_{\hat{Z}}= \frac{m_{Z}^2 + m_{A'}^2  - \sqrt{(m_Z^2 - m_{A'}^2)^2_{} + 4 m_{ZA'}^4 }}{2}\,,\nonumber\\
\hat{A}'^{}_\mu &=& Z_\mu^{} \sin \zeta + A'^{}_\mu \cos \zeta ~~\textrm{with}\nonumber\\
&&m^2_{\hat{A}'}=  \frac{m_{Z}^2 + m_{A'}^2  + \sqrt{(m_Z^2 - m_{A'}^2)^2_{} + 4 m_{ZA'}^4 }}{2}\,,\nonumber\\
&&
\end{eqnarray}
where the rotation angle is determined by
\begin{eqnarray}
\tan 2 \zeta =  \frac{2 m_{ZA'}^2 }{m_{A'}^2 - m_{Z}^2 }\,.
\end{eqnarray}
The $A^{}_\mu$, $\hat{Z}_\mu^{}$ and $\hat{A}'^{}_\mu$ bosons couple to the SM fermions,
\begin{eqnarray}
\mathcal{L}&\supset& \frac{g}{c_W^{}}J_{NC}^\mu  \left(Z_\mu^{}  - \bar{\epsilon}_Z^{} A'^{}_\mu \right)  +  eJ_{em}^{\mu} \left(A_\mu^{} - \bar{\epsilon}_A^{} A'^{}_\mu\right)  \nonumber\\
&=& \left[\frac{g \left(\sin\zeta - \bar{\epsilon}_Z^{}\cos\zeta \right) }{c_W^{}}J_{NC}^\mu - e \bar{\epsilon}_A^{} \cos\zeta J_{em}^\mu  \right]\hat{A}'^{}_\mu\nonumber\\
&&+\left[\frac{g \left(\cos\zeta + \bar{\epsilon}_Z^{}\sin\zeta \right) }{c_W^{}}J_{NC}^\mu + e \bar{\epsilon}_A^{} \sin\zeta J_{em}^\mu  \right]\hat{Z}_\mu^{}\nonumber\\
&&+eJ_{em}^{\mu} A_\mu^{}\,, \end{eqnarray}
with $J_{em,NC}^\mu$ being the SM electromagnetic and neutral currents,
\begin{eqnarray}
J_{em}^\mu &=& -\frac{1}{3}\bar{d}\gamma^\mu_{} d + \frac{2}{3} \bar{u}\gamma^\mu_{} u - \bar{e} \gamma^\mu_{} e \,,\nonumber\\
J_{NC}^\mu &=& \frac{1}{4}\bar{d}\gamma^\mu_{} \left[\left(-1+\frac{4}{3}\sin^2_{}\theta_W^{} \right)+ \gamma_5^{}\right]d \nonumber\\
&&+ \frac{1}{4}\bar{u}\gamma^\mu_{} \left[\left(1-\frac{8}{3}\sin^2_{}\theta_W^{} \right)- \gamma_5^{}\right]u \nonumber\\
&&+ \frac{1}{4}\bar{e} \gamma^\mu_{}\left[\left(-1+4 \sin^2_{}\theta_W^{}\right) + \gamma_5^{}\right] e  \nonumber\\
&&+ \frac{1}{4}\bar{\nu}\gamma^\mu_{} \left[ 1-\gamma_5^{}\right]\nu  \,.
\end{eqnarray}

For a small $\epsilon_Z^{}$, the mass eigenstates $\hat{Z},\hat{A}'$ can well approximate to the $Z,A'$ bosons, and their mass eigenvalues $m_{\hat{Z}, \hat{A}'}^2$ can also be simplified, i.e.
\begin{eqnarray}
\hat{Z}_\mu^{} &\simeq &Z_\mu^{}  ~~\textrm{with}~~m^2_{\hat{Z}}\simeq   \frac{g^2_{} v_\phi^2 }{4 \cos\theta_W^2}\,,\nonumber\\
\hat{A}'^{}_\mu &\simeq &  A'^{}_\mu ~~\textrm{with}~~m^2_{\hat{A}'}\simeq  \frac{g_X^2 v_\chi^2}{4(1-\epsilon_Z^2- \epsilon_A^2)}\,.
\end{eqnarray}
In this limiting case, the $\hat{A}'$ boson only interacts with the neutral currents $J_{NC}^\mu$ at the second order in $\epsilon_Z^{}$. The fashion is similar to the interaction between the $\hat{Z}$ boson and the electromagnetic currents $J^\mu_{em}$. We hence can simply take
\begin{eqnarray}
\mathcal{L}\supset   eJ_{em}^{\mu} A_\mu^{} + \frac{g}{c_W^{}}J_{NC}^\mu \hat{Z}_\mu^{} -\bar{\epsilon}_A^{} e J_{em}^\mu \hat{A}'^{}_\mu + \mathcal{O}(\epsilon^2_{Z})\,.
\end{eqnarray}
In this sense, we can name the $A'$ boson as a dark photon. The experimental constraints and implications on the dark photon $A'$ have been studied in a lot of literatures. (For a recent review, see for example \cite{Fabbrichesi:2020wbt})

Actually, the present model could provide a purely dark photon without any couplings to the neutral currents if the $SU(2)_X^{}\times U(1)_Y^{}$ mixing parameter $\epsilon_B^{}$ and the $SU(2)_X^{}\times SU(2)_L^{}$ mixing parameter $\epsilon_W^{}$ are chosen to be
\begin{eqnarray}
\frac{\epsilon_W^{}}{\epsilon_B^{}}=\tan \theta_W^{} \Rightarrow \epsilon_Z^{} =0\,.
\end{eqnarray}
Since the parameters $\epsilon_{B,W}^{}$ are both induced at loop level, they should be quite small. Therefore, the dark photon $A'$ can be only slightly heavier than the charged dark gauge bosons $X^\pm_\mu$, i.e.
\begin{eqnarray}
m_{A'}^2 - m_{X^\pm_{}}^2 &\simeq &m_X^2 \left( \frac{1}{1-\epsilon_Z^2- \epsilon_A^2}-1\right)\nonumber\\
& \simeq & m_X^2\left(\epsilon_Z^2 + \epsilon_A^2 \right) \ll m_X^2\,.
\end{eqnarray}

\section{Dark matter}

Although the neutral dark gauge boson $X^3_\mu$ now has become the dark photon coupling to the SM electromagnetic and neutral currents, its charged partners $X^{\pm}_{\mu}$ can still keep stable because of the $U(1)_D^{}$ global symmetry. We expect the $X^{\pm}_{\mu}$ bosons to account for the dark matter relic. For this purpose, the annihilations of the $X^{\pm}_{\mu}$ bosons into some light species should arrive at a desired strength. This can be achieved, thanks to the dark Higgs boson $h_\chi^{}$, i.e.
\begin{eqnarray}
\mathcal{L}_{\textrm{Dark}}^{}\supset  \frac{m_X^2}{v_\chi^{}} h_\chi^{} X^{+}_\mu X^{-\mu} +   \frac{m_X^2}{v_\chi^2}h_\chi^{2} X^{+}_\mu X^{-\mu}\,.
\end{eqnarray}
For example, if the dark Higgs boson $h_\chi^{}$ is lighter than the charged dark gauge bosons $X^\pm_\mu$, we can expect the annihilations $X^{+}_{} X^{-}_{} \rightarrow h_\chi^{} h_\chi^{} $. Subsequently, the dark Higgs boson can mostly decay into the mediator scalars including dileptons, diquarks and/or leptoquarks. In the case the dark Higgs boson is too heavy to appear in the final states, it can mediate a $s$-channel annihilation of the charged dark gauge bosons into the mediator scalars. In any of these cases, the mediator scalars eventually can decay into the SM fermion pairs. This means we even can expect a leptophilic dark matter if the mediator is dominated by certain dilepton scalar(s) \cite{gos2009,ghsz2009}. Alternatively, we can resort to the coupling between the dark Higgs boson and the SM Higgs boson for the required annihilations.

The dark $SU(2)_X^{}$ gauge bosons as the dark matter have been studied in other scenario. For example, all of three dark $SU(2)_X^{}$ gauge bosons can keep stable if we do not consider the high dimensional operators (\ref{effective}). The dark $SU(2)_X^{}$ gauge bosons thus can provide three degenerate dark matter particles \cite{hambye2009}. Alternatively, like the 't Hooft-Polyakov monopole model \cite{thooft1974,polyakov1974}, the $SU(2)_X^{}$ gauge symmetry can be spontaneously broken down to a dark $U(1)_X^{}$ gauge symmetry by a real dark Higgs triplet. In consequence, the charged dark gauge boson $X^{\pm}_\mu$ can be a stable dark matter particle while the neutral dark gauge boson $X^3_\mu$ does keep massless and does not couple to the SM electromagnetic and neutral currents \cite{bkp2014}.

\section{Conclusion}

In this paper we have demonstrated an interesting scenario where the dark photon and the dark matter particle can have a nearly degenerate mass spectrum. Specifically we consider the non-abelian kinetic mixing between the dark $SU(2)_X^{}$ gauge group and the SM $SU(2)_L^{}\times U(1)_Y^{}$ gauge groups. Because of these non-abelian kinetic mixing, one of the dark gauge bosons becomes the dark photon, meanwhile, the others keep stable to serve as the dark matter particle. The non-abelian kinetic mixing also makes the dark photon slightly heavier than the dark matter. The quasi-degenerate dark photon and dark matter could be tested if the dark photon and the dark matter are both observed in the future.

\textbf{Acknowledgement}: This work is supported by the Natural Science Foundation of the Higher Education Institutions of Jiangsu Province under grant No.\,22KJB140007.

\end{document}